\documentclass[a4paper,11pt]{article}
\usepackage[dvips]{graphicx}
\usepackage{amssymb}
\usepackage{amsmath}
\usepackage{cite}

\begin{document}

\title{Matter Bounce Loop Quantum Cosmology from $F(R)$ Gravity}
\author{
S.D. Odintsov$^{1,2}$\,\thanks{odintsov@ieec.uab.es}, V.K. Oikonomou$^{3}$\,\thanks{v.k.oikonomou1979@gmail.com; voiko@physics.auth.gr}\\ \\
$^{1).}$Institut de Ciencies de l'Espai (IEEC-CSIC), Campus UAB,\\ 
Torre C5-Par-2a pl, E-08193 Bellaterra, Barcelona, Spain\\\\
$^{2).}$ ICREA, Barcelona, Spain \\\\
$^{3).}$Department of Theoretical Physics, Aristotle University of Thessaloniki,\\
54124 Thessaloniki, Greece
} \maketitle

\begin{abstract}
Using the reconstruction method, we investigate which $F(R)$ theories, with or without the presence of matter fluids, can produce the matter bounce scenario of holonomy corrected Loop Quantum Cosmology. We focus our study in two limits of the cosmic time, the large cosmic time limit and the small cosmic time limit. For the former, we found that, in the presence of non-interacting and non-relativistic matter, the $F(R)$ gravity that reproduces the late time limit of the matter bounce solution is actually the Einstein-Hilbert gravity plus a power law term. In the early time limit, since it corresponds to large spacetime curvatures, assuming that the Jordan frame is described by a general metric that when it is conformally transformed to the Einstein frame, produces an accelerating Friedmann-Robertson-Walker metric, we found explicitly the scalar field dependence on time. After demonstrating that the solution in the Einstein frame is indeed accelerating, we calculate the spectral index derived from the Einstein frame scalar-tensor counterpart theory of the $F(R)$ theory and compare it with the Planck experiment data. In order to implement the resulting picture, we embed the $F(R)$ gravity explicitly in a Loop Quantum Cosmology framework by introducing holonomy corrections to the $F(R)$ gravity. In this way, the resulting inflation picture corresponding to the $F(R)$ gravity can be corrected in order it coincides to some extent with the current experimental data.
\end{abstract}


\section*{Introduction}

One of the most striking experimental results in astrophysics and cosmology was confirmed at the end of the 90's predicting, using standard candles as references, that the universe is expanding but, contrary to the up to then status of expansion, in an accelerating way \cite{riess}. Current experimental research aims to enlighten the universe's evolution from early times to late times \cite{planck,bicep}. The experimentally verified late time acceleration generated an important stream or direction of modern cosmology research, with most of the models and scenarios trying to explain this rather curious and unexpected late time acceleration. Along with the verification of the B-mode power spectrum \cite{bicep}, the main aim of research is to consistently describe early time and late time acceleration of the universe within the same theoretical framework. Moreover, the correct description should in some way describe the various cosmological eras of the universe along with the rather smooth and consistent with experimental data, transition between these eras. 

The modified gravity theories, provide a consistent description of the early time and late time acceleration, with the latter being named dark energy and being described as a negative pressure perfect fluid. This dark energy can be consistently described within the theoretical framework of $F(R)$ modified theories of gravity and related modifications. The bibliography on the subject is vast, but for important papers on this vast research topic, the reader is referred to
\cite{reviews1,reviews2,reviews3,reviews4,reviews5,reviews8,reviews9,
importantpapers1,importantpapers2,importantpapers3,importantpapers4,importantpapers5,
importantpapers6,importantpapers8,importantpapers9,importantpapers10,importantpapers11,
importantpapers12,importantpapers13,importantpapers14,importantpapers15,importantpapers17,
importantpapers18,importantpapers19,importantpapers20} and references therein. It's worth mentioning that the first consistent unified description of early time and late time acceleration in the $F(R)$ theories theoretical framework was done in \cite{sergeinojirimodel}. For alternative theories to $F(R)$ gravities, that can actually describe dark energy, the reader is referred to \cite{capo,capo1,peebles,faraonquin,tsujiintjd}.

Nevertheless, any theory that predicts modifications to Einstein gravity has to be confronted with the astrophysical data. The viability constraints to $F(R)$ theories come from planetary, star formation and local tests (see for example
\cite{reviews1,reviews2,importantpapers4}) and also we should bare in mind that a cosmologically viable $F(R)$ theory must have concordance with the $\mathrm{\Lambda}\mathrm{CDM}$ model \cite{importantpapers3,importantpapers4,importantpapers10,importantpapers15,importantpapers17}. In addition, every $F(R)$ theory is formally equivalent to a Jordan frame scalar-tensor counterpart theory with $\omega$ zero and non-zero potential. This Jordan frame scalar-tensor theory is, by means of a conformal transformation, mathematically equivalent to an Einstein frame scalar-tensor theory, the scalaron of which has to be
classical, in order the stability of the theory is ensured (see relevant work in
\cite{reviews1,reviews2,reviews3,reviews4,reviews5} and related to the subject references therein).

The experimental results of Planck \cite{planck} and BICEP \cite{bicep} have narrowed down the set of inflationary models, excluding models such as chaotic inflation generating power law potentials, exponential potential models and inverse power law models. Interestingly enough, the data set received from the aforementioned experiments seem to are favorably inclined towards $R^2$ gravity. In view of this data, the matter bounce scenario in the context of holonomy corrected Loop Quantum Cosmology (LQC hereafter) \cite{LQC1,LQC2,LQC3,LQC4,LQC5sing,LQC6sing,LQC7sing,LQC7sing1,LQC8,LQC9,LQC10,LQC11,LQC12,LQC13,mbounce1,mbounce2,mbounce3,mbounce4,mbounce5,mbounce6,mbouncersquarefr,mbounce8,mbounce9,mbounce10,mbounce11}, predicts solutions consistent or that can be consistent with the experimental data. For an important stream of reviews and important papers on LQC see \cite{LQC1,LQC2,LQC3,LQC4,LQC5sing,LQC6sing,LQC7sing,LQC7sing1,LQC8,LQC9,LQC10,LQC11,LQC12,LQC13,mbounce1,mbounce2,mbounce3,mbounce4,mbounce5,mbounce6,mbouncersquarefr,mbounce8,mbounce9,mbounce10,mbounce11}. The theoretical framework of LQC is quite appealing since singularities, a rather unwanted feature in every physical theory, are resolved in an elegant way \cite{LQC5sing,LQC6sing,LQC7sing,LQC7sing1}.

One appealing matter bounce scenario \cite{mbounce1,mbounce2,mbounce3,mbounce4,mbounce5,mbounce6,mbouncersquarefr,mbounce8,mbounce9,mbounce10,mbounce11} results if it assumed that the universe is filled with only one scalar field with the simplest scalar potential, leading at early times to matter domination when being in the contracting phase \cite{mbounce10,mbounce11}. In references \cite{mbounce10,mbounce11} an analytic solution of the matter bounce scenario was given, along with a numerical study of the allowed orbits that can be in close agreements with the predictions of the Planck data. In addition, the bounce matter solutions where obtained from an $F(T)$ gravity context in \cite{mbounce4} and also teleparallelism in LQC was studied in \cite{mbounce5}. For a study of $R^2$ modified $F(R)$ gravity, in the context of LQC, see \cite{mbouncersquarefr}.

In view of the interesting properties that are attributed to the cosmological solutions originating from the LQC matter bounce matter theory, we shall use a very well known technique \cite{importantpapers12} in order to reconstruct the $F(R)$ theory that can produce the LQC bounce solutions. There are two reconstruction methods for $F(R)$ gravities, one with the additional use of an auxiliary scalar field \cite{importantpapers3} and the other method does not require any auxiliary field \cite{importantpapers12}. In this paper, by using the latter method \cite{importantpapers12}, we shall investigate which $F(R)$ theories can produce the LQC matter bounce solutions. In addition, apart from searching only a pure $F(R)$ theory that produces the LQC cosmology, we shall also take into account the presence of matter fluids and search for the $F(R)$ theory in this case too. We shall take two limits in our study, the early time limit, describing the inflation era, and the large $t$ limit, more convenient for the description of matter domination period. With regards to the early time limit, we shall find the Einstein frame scalar-tensor counterpart of the Jordan frame $F(R)$ theory taking into account the presence of a relativistic matter fluid (radiation) and we try to make contact with the experimental data of the Planck experiment, focusing on the spectral index value. In order to do so, we assume that we conformally transform a metric from the Jordan frame, that produces a flat Friedmann-Robertson-Walker metric in the Einstein frame. Having found the Einstein frame scalar potential corresponding to the Jordan frame $F(R)$ theory, we demonstrate explicitly that acceleration occurs in the Einstein frame and then we compute the spectral index (it is known that the spectral index for two mathematically-equivalent frames maybe effectively the same, see reference \cite{kaiser}). In order to achieve concordance with the experimental data, we embed the $F(R)$ theory in a LQC framework by introducing holonomy corrections and we qualitatively describe the results. The LQC $F(R)$ theory may yield better results with respect to inflation data.

We have to mention that, apart from the $F(R)$ reconstruction method we shall apply in this article, there is also another equally elegant method of reconstruction that uses instead of the Ricci scalar, the torsion scalar $T$. This is the $F(T)$ method, which can yield also very useful results and very relevant to our analysis. Particularly relevant to our study, is reference \cite{cairev3}, where an actual realization of the matter bounce scenario was extensively studied, with the important study of cosmological perturbations within the context of $F(T)$ theories. We shall perform calculations along similar lines of research but in the context of $F(R)$ gravities.

This paper is organized as follows: In section 1 we briefly present the essentials of $F(R)$ theories. In section 2, after introducing all the necessary information regarding the LQC matter bounce solutions, using the reconstructing technique we search which $F(R)$ gravity produces the LQC bounce solutions. We investigate the problem in the small and large cosmic time $t$ limits. In both cases and in the absence of matter fluids we obtain exact analytical solutions for the pure $F(R)$ gravities. The same applies also in the case matter is present but in the large $t$ limit, with the interesting feature of this case being that when collision-less, non relativistic matter is taken into account, the resulting $F(R)$ gravity is of the form $F(R)=R+AR^{p}$. In section 3, by means of a general conformal transformation we find the Einstein frame scalar-tensor counterpart theory to the early time geometrical $F(R)$ theory and we try to compare the results we get for the spectral index corresponding to this theory, with the experimental results coming from the Planck experiment. In section 4 we embed the $F(R)$ theory in a LQC framework explicitly and we qualitatively describe the general picture of the predicted dynamical equations. The conclusions follow in the end of the paper. Finally, in the appendices A and B we present some details with regards to the calculations we performed in the text.

\section{Essentials of $F(R)$ Gravity}

In order to retain the article self-contained, we review the essential
features of $F(R)$ gravity theories considered in the Jordan frame in the metric
formalism. The reader is referred for detailed analysis on these issues in references \cite{reviews1,reviews2,reviews3,reviews4,reviews5,importantpapers1,
importantpapers2,importantpapers3,importantpapers4,importantpapers5,importantpapers6,importantpapers8,importantpapers9,importantpapers10,importantpapers11,importantpapers12,importantpapers13} and
references therein. 

In this article it shall be presumed that the geometric properties of spacetime, on which the $F(R)$ theories are built upon, are described by a pseudo-Riemannian geometrical background, which locally is a Lorentz metric (the Friedmann-Robertson-Walker metric in our case). In addition, the connection is assumed to be a torsion-less, symmetric, and
metric compatible affine connection, very well known as the so-called Levi-Civita connection. Working on such geometric
backgrounds, the Christoffel symbols are equal to:
\begin{equation}\label{christofell}
\Gamma_{\mu \nu }^k=\frac{1}{2}g^{k\lambda }(\partial_{\mu }g_{\lambda \nu}+\partial_{\nu
}g_{\lambda \mu}-\partial_{\lambda }g_{\mu \nu})
\end{equation} 
and furthermore the Ricci scalar becomes:
\begin{equation}\label{ricciscalar}
R=g^{\mu \nu }(\partial_{\lambda }\Gamma_{\mu \nu }^{\lambda}-\partial_{\nu }\Gamma_{\mu \rho
}^{\rho}-\Gamma_{\sigma \nu }^{\sigma}\Gamma_{\mu \lambda }^{\sigma}+\Gamma_{\mu \rho }^{\rho}g^{\mu
\nu}\Gamma_{\mu \nu }^{\sigma}).
\end{equation}
The four dimensional action of $F(R)$ theories in the Jordan frame is equal to:
\begin{equation}\label{action}
\mathcal{S}=\frac{1}{2\kappa^2}\int \mathrm{d}^4x\sqrt{-g}F(R)+S_m(g_{\mu \nu},\Psi_m),
\end{equation}
In the above relation (\ref{action}) $\kappa$ is related to the gravitational constant $\kappa^2=8\pi G$ and in addition $S_m$ stands for the matter action containing the matter fields $\Psi_m$.

In the metric formalism, the equations of motion are obtained by varying the action (\ref{action}) with respect to the metric $g_{\mu \nu}$, and by doing so we obtain the following equations of motion:
\begin{equation}\label{eqnmotion}
F'(R)R_{\mu \nu}(g)-\frac{1}{2}F(R)g_{\mu \nu}-\nabla_{\mu}\nabla_{\nu}F'(R)+g_{\mu \nu}\square
F'(R)=\kappa^2T_{\mu \nu}.
\end{equation} 
In the above, the prime of the $F(R)$ function denotes differentiation with respect to the Ricci scalar, that is $F'(R)=\partial F(R)/\partial R$ and moreover $T_{\mu \nu}$ is the energy momentum tensor. 

One of the intriguing characteristics of $F(R)$ modified gravity theories is that, what actually makes them modified with respect to Einstein-Hilbert theory of gravity, is that they modify the right hand side of the Einstein equations directly, with the left remaining completely unaltered. So practically speaking, $F(R)$ theories introduce some new form of perfect fluid with purely geometric origin. Of course, this reasoning works at background level: one should bear in mind that this is gravitational $F(R)$ fluid. The obtained in relation (\ref{eqnmotion}) equations of motion for $F(R)$ theories can be cast in the following form:
\begin{align}\label{modifiedeinsteineqns}
R_{\mu \nu}-\frac{1}{2}Rg_{\mu \nu}=\frac{\kappa^2}{F'(R)}\Big{(}T_{\mu
\nu}+\frac{1}{\kappa^2}\Big{(}\frac{F(R)-RF'(R)}{2}g_{\mu \nu}+\nabla_{\mu}\nabla_{\nu}F'(R)-g_{\mu
\nu}\square F'(R)\Big{)}\Big{)}.
\end{align}
Hence, we get an additional contribution to the energy momentum tensor $T_{\mu ,\nu }$, originating from the term:
\begin{equation}\label{newenrgymom}
T^{eff}_{\mu \nu}=\frac{1}{\kappa}\Big{(}\frac{F(R)-RF'(R)}{2}g_{\mu
\nu}+\nabla_{\mu}\nabla_{\nu}F'(R)-g_{\mu \nu}\square F'(R)\Big{)}.
\end{equation}
This new term (\ref{newenrgymom}), absent in Einstein-Hilbert gravity, is what actually explicitly models the dark energy in $F(R)$ theories of modified gravity, and there is where the geometric dark energy terminology stems from. Taking 
the trace of equation (\ref{eqnmotion}), we obtain the following equation:
\begin{equation}\label{traceeqn}
3\square F'(R)+R F'(R)-2F(R)=\kappa^2 T,
\end{equation}
with $T$ the energy momentum tensor's trace $T=g^{\mu \nu}T_{\mu \nu}=-\rho+3P$ and, $\rho_m$ and $P_m$ are the total matter-energy density and pressure respectively. 

Equation (\ref{traceeqn}) reveals another degree of freedom that is present in $F(R)$ theories of gravity, with this degree of freedom described by the function $f'(R)$, commonly known as the scalaron field. The equation of motion for this field is equation (\ref{traceeqn}). Finally, in this paper we shall use a flat Friedmann-Lemaitre-Robertson-Walker (FRW hereafter) spacetime of the following form,
\begin{equation}\label{metricformfrwhjkh}
\mathrm{d}s^2=-\mathrm{d}t^2+a^2(t)\sum_i\mathrm{d}x_i^2
\end{equation}
The Ricci scalar in this metric is equal to:
\begin{equation}\label{ricciscal}
R=6(2H^2+\dot{H}),
\end{equation}
where $H(t)$ stands for the Hubble parameter and the ``dot'' indicates differentiation with respect to time.

\section{Reconstruction of $F(R)$ Gravity from Loop Quantum Cosmology Bounce Solutions}

In principle, it is not an easy task to built a well behaved bounce model and there are some simple reasons for this which we now briefly mention. First to mention is that in order the Hubble rate increases and a bounce occurs, the null energy condition for the matter fields contained in most phenomenological models, that is, the sum of the matter energy density and pressure, has to become negative. These models eventually suffer from ghost instabilities, with the exception of Galileon and ghost condensate models, with the latter two although violating the null energy condition, are free from the pathologies of the usual phenomenological model in a flat FRW universe. Moreover, a phenomenologically correct bounce model should in some way solve the primordial anisotropies increase, which occurs during the contracting phase. With respect to Galileon and ghost condensate models, it is up to date unclear whether these theories result from a more fundamental theory \cite{cairev1}. So considering matter with an equation of state being such, so that a bounce occurs, the anisotropies during the contracting phase could cause BKL oscillations and in effect the universe could collapse in chaotic big crunch. There are two promising ways to address this problem, with the first being to consider ekpyrotic matter \cite{cairev1}, which is characterized by a stiff equation of state and in this way the need for very special initial conditions is avoided \cite{cairev1}. The other scenario the matter bounce inflation scenario, which it's reconstruction we study here. For a comprehensive and informative study on both these issues, the reader is referred to \cite{cairev1}. 

The holonomy corrected Friedmann equation in the context of LQC for a matter dominated universe is given by \cite{mbounce10,mbounce11}:
\begin{equation}\label{holcor1}
H^2=\frac{\rho}{3}\left (1-\frac{\rho}{\rho_c}\right ),{\,}{\,}{\,}\dot{\rho}(t)=-3H\rho(t)
\end{equation}
with the matter-energy density being equal to, 
\begin{equation}\label{rhosol}
\rho (t)=\frac{\rho_c}{\frac{3}{4}t^2+1}
\end{equation}
Solving (\ref{holcor1}) and having in mind (\ref{rhosol}) we obtain the following solutions for the scale factor $a(t)$ and the Hubble parameter $H(t)$ in the matter bounce LQC scenario \cite{mbounce10,mbounce11}:
\begin{equation}\label{holcorrLQCsol}
a(t)=\left (\frac{3}{4}\rho_ct^2+1\right )^{1/3},{\,}{\,}{\,}H(t)=\frac{\frac{1}{2}\rho_ct}{\frac{3}{4}\rho_ct^2+1}
\end{equation}
Having at hand these two solutions, we shall reconstruct the $F(R)$ models that can produce such an expansion history for the universe (for a recent study of bounce cosmology in $F(R)$ gravity, see \cite{sergeibounce}). There are two ways of reconstructing $F(R)$ models that describe a specific cosmological evolutions \cite{importantpapers3,importantpapers12}, with the difference that in the first one an auxiliary scalar field is used to construct the $F(R)$ model. In this paper we shall make use of the second method in which no auxiliary field is used and we briefly present it now. As was explicitly demonstrated in \cite{importantpapers12} every FRW cosmology can be realized by a specific $F(R)$ gravity. The action of $F(R)$ gravity is given by,
\begin{equation}\label{action1dse}
\mathcal{S}=\frac{1}{2\kappa^2}\int \mathrm{d}^4x\sqrt{-g}F(R)+S_m(g_{\mu \nu},\Psi_m),
\end{equation}
and the first FRW equation appearing in relation (\ref{modifiedeinsteineqns}) can be written as:
\begin{equation}\label{frwf1}
-18\left ( 4H(t)^2\dot{H}(t)+H(t)\ddot{H}(t)\right )F''(R)+3\left (H^2(t)+\dot{H}(t) \right )F'(R)-\frac{F(R)}{2}+\kappa^2\rho=0
\end{equation}
with $F'(R)=\frac{\mathrm{d}F(R)}{\mathrm{d}R}$ and the Ricci scalar $R$ given by relation (\ref{ricciscal}) as a function of the time variable. The method developed in \cite{importantpapers12} is based on the introduction of a new variable instead of the cosmological time $t$, the e-folding number $N$, which is related to the scale factor as follows:
\begin{equation}\label{efoldpoar}
e^{-N}=\frac{a_0}{a}
\end{equation} 
Then, the first FRW equation (\ref{frwf1}) can be rewritten in terms of the e-fold parameter $N$,
\begin{align}\label{newfrw1}
& -18\left ( 4H^3(N)H'(N)+H^2(N)(H')^2+H^3(N)H''(N) \right )F''(R)
\\ \notag & +3\left (H^2(N)+H(N)H'(N) \right )F'(R)-\frac{F(R)}{2}+\kappa^2\rho=0
\end{align}
in which case the Hubble parameter is regarded as a function of the e-folds $N$ and the derivatives are defined with respect to $N$ too, that is $H'=\mathrm{d}H/\mathrm{d}N$ and $H''=\mathrm{d}^2H/\mathrm{d}N^2$. Using the function $G(N)=H^2(N)$, equation (\ref{newfrw1}) can be further simplified to the following equation:
 \begin{align}\label{newfrw1modfrom}
& -9G(N(R))\left ( 4G'(N(R))+G''(N(R)) \right )F''(R)
\\ \notag & +\left (3G(N)+\frac{3}{2}G'(N(R)) \right )F'(R)-\frac{F(R)}{2}+\kappa^2\rho=0
\end{align}
with $G'(N)=\mathrm{d}G(N)/\mathrm{d}N$ and $G''(N)=\mathrm{d}^2G(N)/\mathrm{d}N^2$. A key point relation to be used so that $G(N)$ is expressed in terms of the Ricci scalar $R$, in the following way:
\begin{equation}\label{riccinrelat}
R=3G'(N)+12G(N)
\end{equation}
Given the functions $a(t)$ and $H(t)$, by using equations (\ref{efoldpoar}) and (\ref{riccinrelat}), the second order differential equation (\ref{newfrw1modfrom}) can be solved with respect to $F(R)$, so the modified gravity giving rise to the cosmology described by $a(t)$ and $H(t)$ can be reconstructed in an explicit way. In reference \cite{importantpapers12} concrete examples were studied in detail so the reader is referred to it for details. 

In this paper the focus is to pin point which $F(R)$ gravity with or without ordinary matter, can produce the bounce cosmological expansion solutions expressed in terms of the $a(t)$ and $H(t)$ given in relation (\ref{holcor1}), that stem from the holonomy corrected FRW equation, in the context of LQC.  

From relations (\ref{holcor1}), it easily follows that the Hubble parameter can be written in terms of the scale factor as follows,
\begin{equation}\label{hpscf}
H^2=\frac{1}{4}\rho_c^2\left (\frac{4}{3}a^{-3}-a^{-6}\right ) 
\end{equation}
which by using (\ref{efoldpoar}) can be expressed as a function of $N$ and recalling that $G(N)=H^2(N)$, we get,
\begin{equation}\label{gnfunction}
G(N)=\frac{\rho_c^2}{4a_0^3}\left (\frac{4}{3}\rho_ce^{-3N}-\frac{1}{a_0^3}e^{-6N} \right )
\end{equation}
For notational convenience we make the following replacements:
\begin{equation}\label{repl1}
A=\frac{\rho_c^2}{4a_0^3},{\,}{\,}{\,}a=\frac{4\rho_c}{3},{\,}{\,}{\,}b=\frac{1}{a_0^3}
\end{equation}
and by using relations (\ref{riccinrelat}) and (\ref{gnfunction}) we can express the e-fold parameter $N$ as a function of $R$, as follows,
\begin{equation}\label{efoldr}
N=-\frac{1}{3}\ln \left ( \frac{-3aA+\sqrt{9a^2A^2+24RAb}}{12Ab}\right )
\end{equation}
In addition, we shall assume that the matter-energy density appearing in equation (\ref{newfrw1modfrom}) is of the form:
\begin{equation}\label{mattenrgydens}
\rho =\sum_i\rho_{i0}a_0^{-3(1+w_i)}e^{-3(N(R))}
\end{equation}
so by setting $S_i=\rho_{i0}a_0^{-3(1+w_i)}$ and using (\ref{efoldr}), the matter-energy density becomes,
\begin{equation}\label{mattrenergy}
\rho =\sum_i S_i\left ( \frac{-3aA+\sqrt{9a^2A^2+24RAb}}{12Ab}\right )^{1+w_i}
\end{equation}
By making the replacement $x=-3aA+\sqrt{9a^2A^2+24RAb}$, after some calculations, the differential equation (\ref{newfrw1modfrom}) takes the form,
\begin{align}\label{bigdiffgeneral1}
& \left(12\sqrt{3}Ab\right )^2\left (x-12aA \right )\left ( x+3aA\right )x^2\frac{\mathrm{d}^2F(x)}{\mathrm{d}x^2}
\\ \notag & +\Big{(}\left (\frac{6a^2A+\frac{1}{3A}\left ( x+3aA\right )^2-3\left ( x+3aA\right )}{8b} \right )\left ( x+3aA\right )12\sqrt{3}Ab \\ \notag & -\left ( 12\sqrt{3}bA\right )^2x^2\left ( x-12aA\right ) \Big{)}\frac{\mathrm{d}F(x)}{\mathrm{d}x}
\\ \notag & -\left ( x+3aA\right )^2\frac{F(x)}{2}+\sum_iB_{i1}x^{1+w_i}\left ( x+3aA\right )^2=0
\end{align}
where we have set,
\begin{equation}\label{setb1}
B_{i1}=\frac{S_i}{12Ab}
\end{equation}
The differential equation (\ref{bigdiffgeneral1}) is a non-homogeneous generalized Heun equation \cite{mathpaper}, with the difference that the coefficient of the second derivative has $x=0$ as a double root, so it is rather hard to solve it explicitly. It is easy to prove however that no polynomial solutions $F(x)$ exist. In order to see this, let us quote a theorem relevant to this, taken from reference \cite{mathpaper}, that states:

\bigskip

Given the second order differential equation ,
\begin{equation}\label{sectheorem}
\left ( X(x)\frac{\mathrm{d}^2S(x)}{\mathrm{d}x^2}+Y(x)\frac{\mathrm{d}S(x)}{\mathrm{d}x}+Z(x)S(x)=0\right )
\end{equation}
with $X(x)=\sum_{k=0}^4a_kx^k$, $Y(x)=\sum_{k=0}^3b_kx^k$ and $Z(x)=\sum_{k=0}^2c_kx^k$, a degree $n$ polynomial solution of the form $S(x)=\Pi_i(x-x_i)^n$ exists if the following conditions are simultaneously satisfied:
\begin{align}\label{condforpolsol}
&c_2=-n(n-1)a_4-nb_3 \\ \notag &
c_1=-\left (2(n-1)a_4+b_3 \right )\sum_{i=1}^nx_i-n(n-1)a_3-nb_2 \\ \notag &
c_0=-\left ( 2(n-1)a_4+b_3\right )\sum_{i=1}^nx_i^2-2a_4\sum_{i<j}^nx_ix_j \\ \notag &
-\left ( 2(n-1)a_3+b_2\right )\sum_{i=1}^nx_i-n(n-1)a_2-nb_1 
\end{align}
with $x_i$ the $n$ distinct roots of the polynomial solution $S(x)$. In addition, the roots $x_i$ of the polynomial solution $S(x)$, satisfy the Bethe ansatz equations:
\begin{equation}\label{betheansatz}
\sum_{i\neq j}^n\frac{2}{x_i-x_j}+\frac{b_3x_i^3+b_2x_i^2+b_1x_i+b_0}{a_4x_i^4+a_3x_i^3+a_2x_i^2+a_1x_i+a_0},{\,}{\,}{\,}i=1,2,...,n
\end{equation}

\bigskip

Let us consider for simplicity and for the moment, the homogeneous part of the differential equation (\ref{bigdiffgeneral1}) thus disregarding any matter contribution of any form, at least for the moment. For the case at hand, the conditions (\ref{condforpolsol}) are not satisfied, as it can easily be checked by looking the differential equation (\ref{bigdiffgeneral1}). Indeed, in our case, although the first condition might give an integer $n$, $\in$ $N^*$, when the following condition is satisfied:
\begin{equation}\label{condkser}
A^2b^2=\frac{1}{432\times k^2}
\end{equation}
with $k$ any positive integer of our choice, the rest of the conditions are very difficult to satisfy simultaneously. In the Appendix A we provide all the coefficients of the polynomials $X(x)$, $Y(x)$ and $Z(x)$ corresponding to the differential equation (\ref{bigdiffgeneral1}) for the readers convenience.

So we study the problem at hand in the limiting cases when $t$, the cosmological time tends to infinity and in the case $t$ tends to zero.

\subsection{Large $t$ Approximation of LQC Bounce Solution}

We first study the large $t$ limit of the LQC bounce solutions of relation (\ref{holcorrLQCsol}) in order to investigate which $F(R)$ gravity along with some matter content, can generate this late time cosmology. In the large $t$ limit, the scale factor and the Hubble parameter are given by the following relations:
\begin{equation}\label{largetlimitofbsolu}
a(t)=A_1t^{2/3},{\,}{\,}{\,}H(t)=\frac{2}{3t}
\end{equation}
so that $H^2=\Gamma_1a^{-3}$ and where $A_1$ and $\Gamma_1$ stand for,
 \begin{equation}\label{ab1}
 A_1=\left (\frac{3}{4}\rho_c\right )^{1/3},{\,}{\,}{\,}\Gamma_1=\frac{4A_1^{3/2}}{9a_0^3}
\end{equation}
We shall make use the technique we described in the previous section, hence by making use of (\ref{efoldpoar}), the function $G(N)$ is equal to $G(N)=\frac{B_1}{a_0^3}e^{-3N}$, and thereby, solving equation (\ref{riccinrelat}) with respect to $R$, we get:
\begin{equation}\label{nrlarget}
N=-\frac{1}{3}\ln \left ( \frac{R}{\Gamma_1}\right )
\end{equation} 
Substituting $N(R)$ from the above relation, to equation (\ref{newfrw1modfrom}), we obtain the following differential equation,
 \begin{align}\label{diffelargetapprox}
& 3R^2F''(R) -\frac{R}{2}F'(R)-\frac{F(R)}{2}+\sum_iS_i\left ( \frac{R}{\Gamma_1}\right )^{1+w_i}=0
\end{align}
Let us first find which pure $F(R)$ gravity with no content of matter fluids may produce this kind of cosmology, described by the large $t$ limit of the LQC bounce solutions. Without the matter content, the differential equation (\ref{diffelargetapprox}) becomes,
\begin{align}\label{diffelargetapprox1}
& 3R^2F''(R) -\frac{R}{2}F'(R)-\frac{F(R)}{2}=0
\end{align}
which is the Euler second order differential equation, with solutions $f_1(R)$ and $f_2(R)$,
\begin{equation}\label{dgfre}
f_1(R)=R^{\rho_1},{\,}{\,}{\,}f_2(R)=R^{-\rho_2}
\end{equation}
with $\rho_1=27/2$ and $\rho_2=-1/2$. Hence, the pure $F(R)$ theory that generates the cosmology described by relation (\ref{largetlimitofbsolu}), is,
\begin{equation}\label{frgenerlarget}
F(R)=c_1R^{\rho_1}+c_2R^{-\rho_2}
\end{equation}
We now add matter to the $F(R)$ theory so we see which modified gravity theory with matter content can produce (\ref{largetlimitofbsolu}), assuming that matter is described by cold dark matter, so that $w_i=0$. Then, the differential equation (\ref{diffelargetapprox}) becomes,
\begin{align}\label{diffelargetapprox2}
& 3R^2F''(R) -\frac{R}{2}F'(R)-\frac{F(R)}{2}+\frac{S_i}{\Gamma_1}R=0
\end{align}
Suppose the solution to differential equation (\ref{diffelargetapprox2}) is of the form,
\begin{equation}\label{diffsolnonhom}
F(R)=c_1(R)f_1(R)+c_2(R)f_2(R)
\end{equation}
with $f_1(R)$ and $f_2(R)$ being the solutions of the homogeneous differential equation (\ref{diffelargetapprox1}). In order to find $c_1(R)$ and $c_2(R)$ we solve the following system of differential equations:
\begin{align}\label{diffeqnare4}
& f_1(R)c_1'(R)+f_2(R)c_2'(R)=0 \\ \notag &
f_1'(R)c_1'(R)+f_2'(R)c_2'(R)=-\frac{S_i}{\Gamma_1x}
\end{align}
Solving this system, we obtain the following solutions,
\begin{equation}\label{newsolutionsnoneuler}
F(R)=\left ( \frac{c_1S_i-c_2S_i}{\Gamma_1(1-\rho_2)\rho_2} \right )R+\left ( c_2\rho_1-\frac{c_1}{\rho_2(\rho_2-\rho_1+1)}\right )R^{\rho_2+1}
\end{equation}
with $c_1,c_2$ arbitrary constants and $\rho_{1,2}$ defined below equation (\ref{dgfre}). It is quite intriguing that the late time cosmology, that is equivalent to a small $R$ limit, is described by an $F(R)$ function that contains Einstein gravity plus a fractional power of the scalar curvature $R$. Note that we can choose the arbitrary constants in such a way so that the coefficient of the term proportional to $R$ is equal to one, that is:
\begin{equation}\label{approx}
 \frac{c_1S_i-c_2S_i}{\Gamma_1(1-\rho_2)\rho_2}=1
\end{equation}
so that the $F(R)$ action is the Einstein-Hilbert one plus an extra $f(R)$ term. Therefore it is quite intriguing that the late time era for a universe described by the $F(R)$ gravity that reproduces the matter bounce late time cosmological behavior, corresponds to an $F(R)$ gravity with non-relativistic matter, which is the Einstein-Hilbert gravity, plus a power law term.

\subsection{Small $t$ Approximation of LQC Bounce Solution}

In the small $t$ approximation, the scale factor and the Hubble parameter behave in the following way,
\begin{equation}\label{lowtbehav}
a(t)\sim 1+\frac{\rho_ct^2}{4},{\,}{\,}{\,}H(t)\sim \frac{\rho_ct}{2}
\end{equation}
and therefore, by using (\ref{efoldpoar}) and (\ref{riccinrelat}), the e-fold parameter $N$ as a function of $R$ reads,
\begin{equation}\label{efoldr}
N=\frac{1}{3}\ln \left ( \frac{12g_1g_2+ R}{21g_1}\right )
\end{equation}
where we have set $g_1=\rho_ca_0$ and $g_2=\rho_c$. Thereby, the differential equation (\ref{newfrw1}) can be written in terms of the Ricci scalar $R$ and can be written as follows,
\begin{align}\label{ricciscalardiffeqnlarget}
& -9 g_1 (12 g_1 g_2+R) \left(-g_2+\frac{12 g_1 g_2+R}{21 g_1}\right)F''(R)+\left (\frac{9 g_1 g_2}{7}+\frac{5 R}{14}\right )F'(R)\\ \notag &
-\frac{F(R)}{2}+\sum_i B_2(R+12g_1g_2)^{1+w_i}=0
\end{align} 
where $B_2$ stands for,
\begin{equation}\label{b2standsfor}
B_2=\frac{\rho_{i0}a_0^{3(1+w_i)}}{(21)^{1+w_i}} 
\end{equation} 
As we did in the previous large $t$ case, we shall first investigate which pure $F(R)$ gravity (with no matter content) can produce the early time cosmology described by relations (\ref{lowtbehav}). After making the replacement $R=\gamma_1 x$, the differential equation (\ref{ricciscalardiffeqnlarget}) can be written in the following way, 
\begin{align}\label{ricciscalardiffeqnlarget}
\left (x-1\right )\left (x+\frac{4}{3}\right )F''(x)-\frac{45}{6}\left (x+2 \right )F'(x)+\frac{7}{6}F(x)=0
\end{align} 
which can be written as a homogeneous Gauss Hypergeometric equation. In order to see this, the above differential equation is of the form,
\begin{equation}\label{diffeqnnewform}
\left (a_2x^2+b_2x+c_2\right )F''(x)+\left (b_1x+c_1\right )F'(x)+c_0F(x)=0
\end{equation}
with the parameters being,
\begin{equation}\label{parameters}
a_2=1,{\,}{\,}b_2=\frac{1}{3},{\,}{\,}c_2=-\frac{4}{3},{\,}{\,}b_1=-\frac{45}{6},{\,}{\,}c_1=-\frac{45}{3},{\,}{\,}c_0=\frac{7}{6}
\end{equation}
with the polynomial coefficient of $F''(x)$ having roots $\lambda_1=-4/3$ and $\lambda_2=1$. We make the substitution,
\begin{equation}\label{changeofvariable}
z=\frac{x-\lambda_1}{\lambda_2-\lambda_1}
\end{equation}
so that the final form of the differential equation (\ref{ricciscalardiffeqnlarget}) is:
\begin{equation}\label{nobf}
z\left (z-1 \right )F''(z)+\left (Az+B\right )F'(z)+CF(z)=0
\end{equation}
where we have set,
\begin{equation}\label{parmbig}
A=\frac{b_1}{a_2},{\,}{\,}B=\frac{b_1\lambda_1+c_1}{a_2(\lambda_2-\lambda_1)},{\,}{\,}C=\frac{c_0}{a_2}
\end{equation}
The differential equation (\ref{nobf}) has as solution the Gauss Hypergeometric function, $F(z)= F_1(\alpha , \beta , \gamma , z)$, with,
\begin{equation}\label{parafinalnew}
\alpha \beta =\frac{c_0}{a_2},{\,}{\,}\alpha +\beta +1=A,{\,}{\,}\gamma =-B
\end{equation}
Solving the first two equations of relation (\ref{parafinalnew}) with respect to the parameters $\alpha ,\beta$, we obtain a set of two solutions which are,
\begin{equation}\label{albsol}
\alpha =\frac{-51-\sqrt{2433}}{12},{\,}{\,}{\,}\beta =\frac{-51+\sqrt{2433}}{12},{\,}{\,}{\,}\alpha =\frac{-51+\sqrt{2433}}{12},{\,}{\,}{\,}\beta =\frac{-51-\sqrt{2433}}{12}
\end{equation}
It's worth finding an approximation of the Gauss Hypergeometric function in the large $R$ limit, since the small $t$ limit actually describes the inflationary era and we shall need an explicit form of the $F(R)$ action in order to make contact with Planck data of inflation. Recalling that $z=(x-\lambda_1)/(\lambda_2-\lambda_1)$ and $R=\gamma_1 x$, the Gauss Hypergeometric function $F_1(\alpha , \beta , \gamma , z)$ when $R$ goes to infinity can be approximated with the following expression:
\begin{align}\label{finalrelowtexpres}
& F(R)\sim  \frac{\Gamma(\gamma )\left(\frac{\gamma_1}{\lambda_1-\lambda_2}\right)^{-\alpha } \Gamma(-\alpha+\beta) }{\Gamma(\beta) \Gamma(-\alpha+\gamma)}R^{-\alpha}+\frac{\Gamma(\gamma )\left(\frac{\gamma_1}{\lambda_1-\lambda_2}\right)^{-\beta} \Gamma(\alpha-\beta)}{\Gamma(\alpha) \Gamma(-\beta+\gamma)}R^{-\beta}
\end{align}
We can further simplify the above expression by taking into account the values of $\alpha$ and $\beta$ given in relation (\ref{parameters}). By observing these values it is obvious that studying only one of the two cases, automatically provides the solution to the other set of values, since these are symmetric. Taking only the first two values for $\alpha$ and $\beta$, and observing that both $\alpha$ and $\beta$ are negative numbers, since we are studying the large $R$ case, the first term of relation (\ref{finalrelowtexpres}) dominates, so we disregard the second term and the final expression looks like,
\begin{align}\label{finalrelowtexpres1}
& F(R)=A_4 R^{-\alpha}
\end{align}
with $A_4$ being equal to:
\begin{equation}\label{fhequal}
A_4= \frac{\Gamma(\gamma )\left(\frac{\gamma_1}{\lambda_1-\lambda_2}\right)^{-\alpha } \Gamma(-\alpha+\beta) }{\Gamma(\beta) \Gamma(-\alpha+\gamma)}
\end{equation}
This simplification is necessary so we can analytically study the cosmological evolution of the universe affected by $F(R)$ function, during the inflationary period of expansion. However, this may modify the final results as we shall see, therefore a concrete numerical analysis maybe required. In a later section we shall make use of relation (\ref{finalrelowtexpres1}), in order to make contact with Planck data.

\subsubsection{Small $t$ Approximation with Relativistic Matter}

We now study the case in which, apart from the $F(R)$ modified gravity in the small $t$ limit, we also take into account the presence of a relativistic matter fluid ($w=1/3$). Then, the differential equation (\ref{nobf}) is written as follows:
\begin{equation}\label{nobf1}
z\left (z-1 \right )F''(z)+\left (Az+B\right )F'(z)+CF(z)=21B_2\left ( z(\lambda_2-\lambda_1)+\lambda_1+\frac{4}{3} \right )^{4/3}
\end{equation}
where $A,B,C,B_2$ are given in relations (\ref{b2standsfor}), (\ref{parameters}) and (\ref{parmbig}). In order to find a solution to this equation, which we shall denote $y_2(z)$, we shall make use of the solution we found corresponding to the non-homogeneous case, which we denote $y_1(z)$ and is given by the Gauss hypergeometric function, that is:
\begin{equation}\label{djffff}
y_1(z)=F(z)=F_1(\alpha , \beta , \gamma , z)
\end{equation} 
Then we can reduce the order of the non-homogeneous differential equation (\ref{nobf1}) using well know techniques. For the shake of notational simplicity we shall introduce some new notation. We set:
\begin{align}\label{notationnewsimple}
& P(z)=z(1-z),{\,}{\,}{\,}Q(z)=-(Az+B) \\ \notag &
G(z)=21B_1\left (z(\lambda_2-\lambda_1+\frac{4}{3}+\lambda_1 \right )^{4/3}
\end{align}
Then, the solution of (\ref{nobf1}) can be written $y_2(z)=y_1(z)v(z)$ with $u(z)=v'(z)$ and $u(z)$ being equal to:
\begin{equation}\label{finalieseqn}
u(z)=e^{-\int A_1(z)\mathrm{d}z}\left (\int B_1(z)e^{\int A_1(z)} +c\right )
\end{equation}
with $A_1(z)$, $B_1(z)$ are defined to be:
\begin{equation}\label{fhf}
A_1(z)=2\frac{y_1'(z)}{y_1(z)}+\frac{Q(z)}{P(z)},{\,}{\,}{\,}B_1(z)=\frac{G(z)}{P(z)y_1(z)}
\end{equation}
The final solution of the non-homogeneous differential equation (\ref{nobf1}) is of the following form:
\begin{equation}\label{solform}
y_2(z)=y_1(z)v(z)
\end{equation}
with $v(z)$ being equal to:
\begin{equation}\label{finalslolexpressed}
v(z)=\int u(z) \mathrm{d}z+C
\end{equation}
In order to compute the solution, we shall take the limit $z\rightarrow \infty$ which is equivalent to the large $R$ limit. By omitting some tedious calculations, the resulting expression for the solution of the non-homogeneous equation $F(z)$, expressed as a function of the Ricci scalar $R$, is given by the following expression:
\begin{align}\label{fgdds}
& F(R)=\frac{12\text{  }B_1 \Gamma\left(\frac{4}{3}+\alpha \right) \Gamma(\alpha -\beta )^2 \Gamma(\beta ) \Gamma(\gamma ) \Gamma(-\alpha +\gamma ) \lambda _1 \left(\frac{\gamma _1}{-\lambda _1+\lambda _2}\right)^{\frac{4}{3}-A+2 \alpha -2 \beta } \left(-\lambda _1+\lambda _2\right)^{1/3}}{\left(\frac{1}{3}-A+B+\alpha \right) 
(1+3 \alpha ) \left(\frac{4}{3}-A+2 \alpha -\beta \right) \Gamma(\alpha )^2 \Gamma\left(\frac{1}{3}+\alpha \right) \Gamma(-\alpha +\beta ) \Gamma(-\beta +\gamma )^2}R^{\frac{4}{3}-A+2 \alpha -2 \beta } \\ \notag & +\frac{12\text{  }B_1 \Gamma\left(\frac{4}{3}+\alpha \right) \Gamma(\alpha -\beta ) \Gamma(\gamma ) \lambda _1 \left(\frac{\gamma _1}{-\lambda _1+\lambda _2}\right)^{\frac{4}{3}-A+\alpha -\beta } \left(-\lambda _1+\lambda _2\right)^{1/3}}{\left(\frac{1}{3}-A+B+\alpha \right) (1+3 \alpha ) \left(\frac{4}{3}-A+2 \alpha -\beta \right) \Gamma(\alpha ) \Gamma\left(\frac{1}{3}+\alpha \right) \Gamma(-\beta +\gamma )}R ^{\frac{4}{3} -A+\alpha -\beta }
\end{align}
The details of the approximations we made can be found in Appendix B. Taking into account the values of the parameters defined in relations (\ref{parameters}) and (\ref{albsol}), we can approximate the $F(R)$ function (\ref{fgdds}) as follows,
\begin{align}\label{inflapprxo}
F(R)=A_4R^{\delta}
\end{align}
with the parameter $\delta $ being equal to:
\begin{equation}\label{deltapar}
\delta = \frac{4}{3}-A+\alpha -\beta  
\end{equation}
and the coefficient $A_4$ defined as:
\begin{equation}\label{a4coeff}
A_4=\frac{12 \Gamma\left(\frac{4}{3}+\alpha \right) \Gamma(\alpha -\beta ) \Gamma(\gamma ) B_1 \lambda _1 \left(\frac{\gamma _1}{-\lambda _1+\lambda _2}\right)^{\frac{4}{3}-A+\alpha -\beta } \left(-\lambda _1+\lambda _2\right)^{1/3}}{\left(\frac{1}{3}-A+B+\alpha \right) (1+3 \alpha ) \left(\frac{4}{3}-A+2 \alpha -\beta \right) \Gamma(\alpha ) \Gamma\left(\frac{1}{3}+\alpha \right) \Gamma(-\beta +\gamma )}
\end{equation}
Notice that many viable $F(R)$ theories maybe approximated at large or small cosmic time regime, as some power-law models \cite{reviews4}, (relation II.68 ), i.e. the models we precisely obtained here. Of course, it is rather hard, to analytically reconstruct such non-linear functions, this can be done only numerically. Notice that reconstructed models can have relation with viable gravities based on the Khoury chameleon scenario \cite{khoury}.

In the next section we shall attempt to make contact with the Planck data regarding inflation for the $F(R)$ function corresponding to the pure modified gravity theory, in the absence of the matter fluid. We have to note that in order to obtain the results in an analytically described closed form, we shall use the approximation (\ref{finalrelowtexpres}) for the pure $F(R)$ function, so we expect some deviations from the experimental data.

\section{Einstein Frame Inflationary Potential and Comparison with Planck Data}

In this section we shall study inflation by studying the scalar-tensor theory corresponding to the $F(R)$ function appearing in relation (\ref{finalrelowtexpres1}) (for a general description of $F(R)$ inflation, see \cite{sergeiinflnew}). A basic question arises when someone wants to consider inflation issues with regards to the Einstein frame having to do with the de-Sitter (in general an acceleration producing metric) metric in the Einstein frame. In order inflation occurs in the Einstein frame, it is compelling to have a de-Sitter (accelerating) FRW metric describing the evolution of the universe in that frame. Now it would be appealing that the bounce FRW metric under a specific conformal transformation produces the de-Sitter (accelerating) metric in the Einstein frame. This actually occurs in the present case as we shall demonstrate shortly. Then, there appears the beautiful picture: what looks as bounce in one frame, maybe inflation in the conformally related frame, or to put it in a different way, it could be that, what looks as bounce in one frame, may look as inflation in mathematically-equivalent frame. We give some arguments to support this conjecture below.

Our strategy is to assume that we start at the Jordan frame, with the universe being described by an $F(R)$ gravity in a metric background with metric tensor that when it is conformally transformed to the Einstein frame, it produces the de-Sitter (in general accelerating) expansion in that frame. In principle, the Jordan frame metric is a solution of the Einstein equations in the Jordan frame, but the particular form of it is of no importance for the moment. Notice that in this approach the Jordan frame is the unphysical frame. So conformally transforming this metric we obtain a de-Sitter (accelerating) metric in the Einstein frame. Then, by solving the slow-roll inflation equations, we may obtain the solution with respect to the scalar field that describes inflation. It is exactly that field that enters in the conformal transformation from the Jordan frame. Having at hand the scalar field, we may explicitly check whether this transformation produces a matter bounce FRW universe in the Jordan frame, or equivalently if we use this slow roll inflation producing scalar field in the conformal transformation, then can the Jordan frame matter bounce metric produce some accelerating cosmology?

In order to study inflation, we shall express the Jordan frame theory in terms of the Einstein frame, by means of a conformal transformation. We assume that the Jordan frame metric is such that, when conformally transformed to the Einstein frame, it produces an accelerating or de-Sitter expansion to the Einstein frame. This technique is quite well known and can be found in most reviews articles, see for example \cite{reviews1,reviews2} and also \cite{sergeistarobinsky} for a concrete similar analysis to the one we perform here. Starting from the action (\ref{action1dse}), neglecting matter fields and introducing the auxiliary field $A$, the action (\ref{action1dse}) can be written as follows:
\begin{equation}\label{action1dse111}
\mathcal{S}=\frac{1}{2\kappa^2}\int \mathrm{d}^4x\sqrt{-g}\left ( F'(A)(R-A)+F(A) \right ),
\end{equation}
It can be easily checked that variation with respect to $A$ yields the solution $A=R$, a fact that indicates the equivalence of the two actions (\ref{action1dse}) and (\ref{action1dse111}). The canonical transformation that will relate the Jordan frame $F(R)$ theory to the Einstein frame scalar tensor theory is defined by means of the following functional relation,
\begin{equation}\label{can}
\sigma =-\sqrt{\frac{3}{2k^2}}\ln (F'(A))
\end{equation}
where we introduced the scalaron field $\sigma$, which will be a scalar degree of freedom in the Einstein frame. Using the conformal transformation of the Jordan frame metric $g_{\mu \nu }$,
\begin{equation}\label{conftransmetr}
\tilde{g}_{\mu \nu}=e^{-\sigma }g_{\mu \nu }
\end{equation}
where the ''tilde'' denotes the Einstein frame metric, we obtain the Einstein frame action,
\begin{align}\label{einsteinframeaction}
& \mathcal{\tilde{S}}=\int \mathrm{d}^4x\sqrt{-\tilde{g}}\left ( \frac{\tilde{R}}{2k^2}-\frac{1}{2}\left (\frac{F''(A)}{F'(A)}\right )^2\tilde{g}^{\mu \nu }\partial_{\mu }A\partial_{\nu }A -\frac{1}{2k^2}\left ( \frac{A}{F'(A)}-\frac{F(A)}{F'(A)^2}\right ) \right ) \\ \notag &
= \int \mathrm{d}^4x\sqrt{-\tilde{g}}\left ( \frac{\tilde{R}}{2k^2}-\frac{1}{2}\tilde{g}^{\mu \nu }\partial_{\mu }\sigma\partial_{\nu }\sigma -V(\sigma )\right )
\end{align}
with $V(\sigma )$, where again the ''tilde'' denotes quantities in the Einstein frame. The potential as a function of $\sigma $ is,
\begin{align}\label{potentialvsigma}
V(\sigma )=\frac{A}{F'(A)}-\frac{F(A)}{F'(A)^2}=\frac{1}{2k^2}\left ( e^{\sqrt{2k^2/3}\sigma }R\left (e^{-\sqrt{2k^2/3}\sigma} \right )- e^{2\sqrt{2k^2/3}\sigma }F\left [ R\left (e^{-\sqrt{2k^2/3}\sigma} \right ) \right ]\right )
\end{align}
with the function $R\left (e^{-\sqrt{2k^2/3}\sigma} \right )$ being the solution of equation (\ref{can}), with respect to $A$, with $A=R$. Having at hand the Einstein frame action, we can easily obtain the essentials of the inflation framework, in order to make contact with experimental data of Planck. The energy density and the pressure of the scalar field are given by,
\begin{equation}\label{ingflatrg}
\rho_{\sigma }=\frac{\dot{\sigma}^2}{2}+V(\sigma ),{\,}{\,}{\,}p_{\sigma }=\frac{\dot{\sigma}^2}{2}-V(\sigma )
\end{equation}
with the ''dot'' indicating as always, derivative with respect to the cosmological time. The Friedmann equations in the presence of the $\sigma $ field in the Einstein frame are equal to:
\begin{equation}\label{einsteinfrweqns}
\frac{3H^2}{k^2}=\frac{\dot{\sigma}^2}{2}+V(\sigma ),{\,}{\,}{\,}-\frac{1}{k^2}\left ( 2\dot{H}+3H^2\right )=\frac{\dot{\sigma}^2}{2}-V(\sigma )
\end{equation}
while the energy conservation law yields a second order equation for the inflaton field,
\begin{equation}\label{secordereqns}
\ddot{\sigma}+3H\dot{\sigma}=-V'(\sigma )
\end{equation}
Here the prime denotes differentiation with respect to the inflaton field $\sigma $. The acceleration of the inflationary universe can be written in terms of the slow roll parameter $\epsilon$ as follows,
\begin{equation}\label{acclereps}
\frac{\ddot{a}}{a}=H^2(1-\epsilon)
\end{equation}
with $\epsilon$ being equal to,
\begin{equation}\label{fvareps}
\epsilon =-\frac{\dot{H}}{H^2}=\frac{k^2\dot{\sigma}^2}{2H^2}
\end{equation}
There is also another slow roll parameter denoted as $\eta$ and is defined to be,
\begin{equation}\label{slowrolleta}
\eta =\epsilon-\frac{\dot{\epsilon }}{2\epsilon H}=-\frac{\ddot{\sigma }}{H\dot{\sigma }}
\end{equation}
A fundamental requirement for the existence of a consistent inflationary era, is that the universe evolves into a quasi de-Sitter space, in which state it remains for a sufficient period of time. In order this to be true, the quantity $\dot{H}$ and consequently the slow roll parameters must be very small. Thereby, the kinetic energy of the field during the inflationary era has to be small, a fact that is expressed quantitatively by the following requirement,
\begin{equation}\label{lowinflareq}
\dot{\sigma}^2\ll V(\sigma)
\end{equation}
and therefore the slow roll parameters can be expressed as functions of the inflaton potential in the following way,
\begin{equation}\label{slowrollinflpotent}
\epsilon =\frac{1}{2k^2}\left (\frac{V'(\sigma)}{V(\sigma)}\right ),{\,}{\,}{\,}\eta =\frac{1}{k^2}\left ( \frac{V''(\sigma )}{V(\sigma )} \right )
\end{equation}
The inflationary period ends when $\epsilon ,\eta\sim 1$. Finally, in order to make contact with the Planck data, we will use the spectral index corresponding to the slow roll inflationary period,
\begin{equation}\label{slowrollspectralindex}
n_s=1-6\epsilon +2\eta
\end{equation}
with the last Planck data constraining the allowed values to be (see \cite{sergeistarobinsky}),
\begin{equation}\label{contraintplanck}
n_s=0.9603\pm 0.0073
\end{equation}
After this brief review of the inflation essentials (for more details consult \cite{mukhanov}), we proceed to calculate the potential in the Einstein frame corresponding to the $F(R)$ function (\ref{finalrelowtexpres1}). The solution of equation (\ref{can}) for the $F(R)$ function (\ref{finalrelowtexpres1}) is,
\begin{equation}\label{solfrocan}
R=e^{\frac{\sqrt{\frac{2}{3}k^2} \sigma }{1+\alpha }} \left(\alpha  A_4\right)^{\frac{1}{1+\alpha }}
\end{equation}
and correspondingly, the potential $V(\sigma )$ can easily be calculated using relation (\ref{potentialvsigma}), and is equal to,
\begin{equation}\label{dpanve}
V(\sigma )=\frac{e^{\sqrt{\frac{2}{3}k^2} \sigma } \left(e^{\frac{\sqrt{\frac{2}{3}} \sqrt{k^2} \sigma }{1+\alpha }} \left(\alpha  A_4\right)^{\frac{1}{1+\alpha }}\right)^{-\alpha } \left(-A_4+\left(e^{\frac{\sqrt{\frac{2}{3}} \sqrt{k^2} \sigma }{1+\alpha }} \left(\alpha  A_4\right)^{\frac{1}{1+\alpha }}\right)^{1+\alpha }\right)}{2 k^2}
\end{equation}
Before we proceed in calculating the slow roll parameters for inflation, we have to show that the expansion in the Einstein frame is indeed accelerating. Solving the pair of the following coupled differential equations, valid in the slow roll limit,
\begin{equation}\label{slowrolldiff}
\frac{3H^2}{k^2}\simeq V(\sigma ),{\,}{\,}{\,}3H\dot{\sigma}\simeq -V'(\sigma )
\end{equation}
with the potential being equal to (\ref{dpanve}), the slow roll limit solution of equations (\ref{slowrolldiff}) is equal to,
\begin{equation}\label{slowrollsolution}
\sigma (t)\simeq -\frac{1}{\gamma_1 }\ln \left ( \gamma_1\mathcal{B}t+t_0\right )
\end{equation}
with the parameters $\gamma_1$ and $\mathcal{B}$ being equal to,
\begin{equation}\label{paramegbeta}
\gamma_1=\sqrt{\frac{2k^2}{3}}\frac{1}{1-\delta},{\,}{\,}{\,}\mathcal{B}=\frac{k(\delta-2)(\delta A_4)^{\frac{1}{2(1-\delta)}}}{3k^2(\delta-1)}
\end{equation}
Consequently we have,
\begin{align}\label{handhdot}
& H(t)\simeq \mathcal{G}\left ( \mathcal{B}\gamma_1 t+t_0\right )^{\frac{2-\delta}{2(\delta-1)}},\\ \notag &
\dot{H}+H^2\simeq \mathcal{G}^2\left ( \mathcal{B}\gamma_1 t+t_0\right )^{\frac{2-\delta}{(\delta-1)}}+\frac{\mathcal{G}\mathcal{B}\gamma_1\left ( \mathcal{B}\gamma_1 t+t_0\right )^{\frac{4-3\delta}{2(\delta-1)}}}{2(\delta-1)}
\end{align}
with $\mathcal{G}$ being equal to,
\begin{equation}\label{mathcalg}
\mathcal{G}=\sqrt{(\delta A_4)^{\frac{1}{1-\delta}}-A_4(\delta A_4)^{\frac{\delta }{1-\delta}}}
\end{equation}
Notice that $\frac{2-\delta}{2(\delta-1)}<0$. In addition, since the following inequalities hold always true,
\begin{equation}\label{ineq}
\mathcal{G}^2>0,{\,}{\,}{\,}\frac{\mathcal{B}\gamma_1\mathcal{G}}{2(-1+\delta )}>0
\end{equation}
the expansion is accelerating, so inflation actually occurs. Now our task is to see to what extend the inflationary solutions in the Einstein frame, agree with the available experimental data. Using relation (\ref{slowrollinflpotent}), we can calculate the slow roll parameters $\epsilon$ and $\eta$, in the slow roll limit of inflation, with $\epsilon $ being equal to,
\begin{equation}\label{epslinol1}
\epsilon =\frac{\left((1+\alpha ) A_4-(2+\alpha ) \left(e^{\frac{\sqrt{\frac{2}{3}} \sqrt{k^2} \sigma }{1+\alpha }} \left(\alpha  A_4\right)^{\frac{1}{1+\alpha }}\right)^{1+\alpha }\right)^2}{3 (1+\alpha )^2 \left(A_4-\left(e^{\frac{\sqrt{\frac{2}{3}} \sqrt{k^2} \sigma }{1+\alpha }} \left(\alpha  A_4\right)^{\frac{1}{1+\alpha }}\right)^{1+\alpha }\right)^2}
\end{equation}
while $\eta$ is equal to,
\begin{equation}\label{etaeqns}
\eta= \frac{-2 (1+\alpha )^2 A_4+2 (2+\alpha )^2 \left(e^{\frac{\sqrt{\frac{2}{3}} \sqrt{k^2} \sigma }{1+\alpha }} \left(\alpha  A_4\right)^{\frac{1}{1+\alpha }}\right)^{1+\alpha }}{3 (1+\alpha )^2 \left(-A_4+\left(e^{\frac{\sqrt{\frac{2}{3}} \sqrt{k^2} \sigma }{1+\alpha }} \left(\alpha  A_4\right)^{\frac{1}{1+\alpha }}\right)^{1+\alpha }\right)}
\end{equation}
In order to be concise with the slow roll approximation, we have to express relations $\epsilon$ and $\eta $ in the $\sigma \rightarrow -\infty$ limit (which correspond to the large curvature limit, see relation (\ref{can})). Recalling that from relation (\ref{albsol}) that $ \alpha $ is negative, then relations (\ref{epslinol1}) and (\ref{etaeqns}) in the $\sigma \rightarrow  -\infty$ limit are approximated to be,
\begin{equation}\label{approxepseta}
\epsilon \simeq \frac{1}{3},{\,}{\,}{\,}\eta \simeq \frac{2\left((2+\alpha )^2-3-2\alpha \right)}{3(1+\alpha )^2}
\end{equation}
so using the numerical value of the parameter $\alpha \simeq-8.3604$, the spectral index is approximately $n_s\simeq 0.33$. Hence compared to the experimental data, we can see that the model described by the $F(R)$ function of relation (\ref{inflapprxo}) does not provide a perfect fit to the Planck experimental data, with regards to the spectral index value. However the model (\ref{inflapprxo}) is just an approximation of the full model given in relation (\ref{fgdds}) so one should take into account the whole $F(R)$ model of relation (\ref{fgdds}). However, it would be rather difficult to solve analytically equation (\ref{can}) when the complete $F(R)$ model of relation (\ref{fgdds}) is taken into account. Instead, we can further implement the $F(R)$ gravity results by embedding the theory directly in a LQC framework by introducing holonomy corrected $F(R)$ gravity. We briefly sketch out how this would work in the next section, deferring the reader to a future detailed  work on these issues. Let us note that in the case at hand, the universe expands in eternal inflationary way. However the ending of the eternal inflation process \cite{guth} may be caused by curvature perturbations, like in the primordial de Sitter universe emerging due to the conformal anomaly \cite{starobinsky,sergeitrace} or owing to instability of the reconstructed $F(R)$ gravity which causes the curvature perturbations. We shall address the latter issue in a future work.

Having the scalar field solution (\ref{slowrollsolution}) which guarantees a slow roll inflation in the Einstein frame, we can explicitly conformal transform the bounce matter FRW metric,
\begin{equation}\label{bouncefrwmetric}
\mathrm{d}s^2=-\mathrm{d}t^2+\left (\frac{3}{4}\rho_ct^2+1\right )^{2/3}\sum_i\mathrm{d}x_i^2
\end{equation}
in order to see if it produces an inflationary-accelerating solution in the Einstein frame. Notice that the following investigation is not related to the Einstein frame accelerating cosmology we found previously, since in that case we started from a general metric that, when conformally transformed, produces an FRW accelerating slow roll cosmology in the Einstein frame. Here we are just interested to see if the matter bounce metric can be conformally related to an accelerating metric, with the choice of the conformal transformation being determined by the slow roll inflationary solution in the Einstein frame, namely equation (\ref{slowrollsolution}).

Performing the conformal transformation of the bounce metric (\ref{bouncefrwmetric}), using $\sigma (t)$ defined in (\ref{slowrollsolution}), we get the following metric,
\begin{equation}\label{nondesittermetric}
\mathrm{d}s^2=-\left ( \mathcal{B}\gamma_1 t+t_0\right)^{-\frac{1}{\gamma_1}}\mathrm{d}t^2+\frac{\left (\frac{3}{4}\rho_ct^2+1\right )^{2/3}}{\left ( \mathcal{B}\gamma_1 t+t_0\right)^{\frac{1}{\gamma_1}}}\sum_i\mathrm{d}x_i^2
\end{equation}
Performing the transformation,
\begin{equation}\label{transfro123}
t\rightarrow \frac{1}{\mathcal{B\gamma_1}}(\mathcal{B}\gamma_1 t+t_0)^{\frac{4\gamma_1-1}{2\gamma_1}}
\end{equation}
the metric becomes,
\begin{equation}\label{nondesittermetrictransformed}
\mathrm{d}s^2=-\mathrm{d}t^2+\frac{(\frac{3\rho_c(\mathcal{B}\gamma_1t^{2\gamma_1}{4\gamma_1-1}-t_0)}{\mathcal{B}\gamma_1})^{1/3}}{( \mathcal{B}\gamma_1)^{-\frac{2}{4\gamma_1-1}}t^{\frac{-1}{4\gamma_1-1}}}\sum_i\mathrm{d}x_i^2
\end{equation}
which gives the following two results, with respect to Hubble parameter,
\begin{align}\label{handhdotaccel}
& H(t)\simeq \frac{3+\gamma _1 \left(2-\frac{2 t_0}{t_0-B t^{\frac{\gamma _1}{2 t_0}} \gamma _1}\right)}{12 t t_0},\\ \notag &
\dot{H}+H^2\simeq \frac{9 t_0^2 \left(1-4 \gamma _1\right)+B^2 t^{\frac{\gamma _1}{t_0}} \gamma _1^2 \left(3+2 \gamma _1\right) \left(3-12 t_0+2 \gamma _1\right)}{144 t^2 t_0^2 \left(t_0-B t^{\frac{\gamma _1}{2 t_0}} \gamma _1\right)^2}\\ \notag &
-\frac{6 B t^{\frac{\gamma _1}{2 t_0}} t_0 \gamma _1 \left(3+2 \gamma _1 \left(1+\gamma _1\right)-4 t_0 \left(3+\gamma _1\right)\right)}{144 t^2 \gamma _1^2 \left(t_0-B t^{\frac{\gamma _1}{2 t_0}} \gamma _1\right)^2}
\end{align}
which, owing to the fact that $\mathcal{B}<0$ and for the appropriate choice of $t_0$, in order $3-12 t_0+2 \gamma _1>0$ can give an accelerating expansion. Therefore we have shown that the bounce matter FRW metric in the Jordan frame can be conformally related to an accelerating metric in the Einstein frame, by choosing the scalar field appearing in the conformal transformation, to be the solution of the slow roll inflation in the Einstein frame. In principle however, it is rather difficult to achieve the conditions that ensure acceleration in the Einstein frame.

\section{Loop Quantum Cosmological Extension of $F(R)$ Gravity}

So far our aim was to reproduce the matter bounce evolution of the universe in terms of the scale factor and the Hubble parameter. Therefore, there was no direct use of the LQC holonomy corrections, directly in the $F(R)$ gravity. It is however very well known that holonomy corrections can be introduced in the $F(R)$ gravity Lagrangian directly \cite{mbounce5,mbouncersquarefr,holonomyfr}. Such a modification of the $F(R)$ gravity Lagrangian, can in principle have very appealing results in the $F(R)$ theory predictions, such as avoidance of early time curvature singularities or even refinements of the inflationary period predictions of the $F(R)$ theory (see reference \cite{mbouncersquarefr} for the study of $R+aR^2$ gravity). In view of these appealing properties that the holonomy corrected $F(R)$ gravity offers, we shall briefly study here the Einstein frame implications of the holonomy corrected $F(R)$ gravity appearing in relation (\ref{inflapprxo}). As we shall demonstrate shortly, certain issues arise that renders this study a kind of intriguing issue which deserves a more detailed study than the one we shall briefly present here, just to sketch out the problem. 

We start off by describing the general framework of holonomy LQC $F(R)$ gravity (for details consult \cite{mbounce5,mbouncersquarefr,holonomyfr}). We assume a flat FRW metric in order to simplify the equations and also work in the Einstein frame, because the Jordan frame problem might be particularly difficult. For the details on the Jordan frame problem see \cite{mbouncersquarefr}. Denoting with a tilde all the Einstein frame quantities, the classical variable $\tilde{\beta}\equiv \gamma \tilde{H}$ and the volume $\tilde{V}=\tilde{a}^3$ are canonically conjugated variables with the Poisson bracket being equal to $\{\tilde{\beta} ,\tilde{V}\}=\gamma/2$, with $\tilde{H}$ and $\gamma $, being the Hubble law parameter and the Barbiero-Immirzi parameter. Owing to the fact that we are working on a discrete space, the state space of vectors belonging to the corresponding Hilbert space consists of periodic functions. In addition, the operator $\beta $ is not a well defined quantum operator and holonomy corrections have to be directly introduced into the Hamiltonian in order to have well defined operators (this non well-defined property comes from the Hamiltonian constraint in which $\beta$ enters directly). Introducing the holonomy corrections (for details on this look \cite{mbouncersquarefr}), the holonomy corrected FRW equation in the Einstein frame is,
\begin{equation}\label{einframeholcorr}
\tilde{H}^2=\frac{1}{3}\tilde{\rho}\left ( 1-\frac{\tilde{\rho}}{\tilde{\rho_c}}\right )
\end{equation}
with $\tilde{\rho}_c$ the critical density in the Einstein frame. As pointed out in \cite{mbouncersquarefr}, equation (\ref{einframeholcorr}) describes an ellipse in the $(\tilde{H},\tilde{\rho })$ plane, with the dynamics of the universe being very simple to describe, since the universe moves clockwise from a contracting phase to an expanding phase, beginning and ending at the critical point $(0,0)$ and bouncing off only once at $(0,\tilde{\rho })$. Let us proceed to make contact with the particular holonomy corrected $F(R)$ gravity at hand, and in order to do so, we shall need the Einstein frame version of the $F(R)$, which we studied in the previous section. Particularly we shall need the potential (\ref{dpanve}), which can be obtained by making the conformal transformation from the Jordan frame (\ref{can}). By doing so, the equation describing the scalar field evolution in the Einstein frame is,
\begin{equation}\label{scalrevol}
\ddot{\sigma}+3\tilde{H}\dot{\sigma }+\frac{\partial V(\sigma)}{\partial{ \sigma}}=0,
\end{equation}
with the potential being the one appearing in equation (\ref{dpanve}). The physics described by the holonomy corrected LQC can be resumed in the following scenario, in which the universe starts in the contracting phase with zero energy, the energy increases up to the energy equal to the Einstein frame critical density and then it bounces of entering in the accelerating phase. In our case, by performing the change of variable $\sqrt{\frac{2k^2}{3}}\psi =\ln \sigma$, the evolution equation for the scalar field $\psi$ reads,
\begin{align}\label{psievol}
& \ddot{\psi}\psi+3\tilde{H}\dot{\psi }\psi+
\\ \notag & +\frac{1}{\sqrt{6}(\delta-1)k}\Big{(}(\delta-2)(\delta A_4)^{\frac{1}{1-\delta}}\psi^{n+1}-2(\delta-1)(\delta A_4)^{\frac{\delta}{1-\delta}}\psi^{n} \Big{)}
\end{align}
with ''$n$'' being related to the parameter $\delta$ as follows,
\begin{equation}\label{deltan}
n=\frac{2\delta-3}{\delta-1}
\end{equation}
As shown in reference \cite{mbouncersquarefr}, the equation (\ref{psievol}) has some symmetries, which for the case at hand are particularly simple to see. Actually the orbits of the dynamical system at hand, being described in phase space by $(\dot{\psi},\psi )$, have a symmetry axis, the $\dot{\psi}$ axis, since equation (\ref{psievol}) remains invariant under the transformation 
\begin{equation}\label{ttransfrom}
\tilde{t}\rightarrow -\tilde{t},{\,}{\,}{\,}\tilde{H}\rightarrow -\tilde{H}
\end{equation}
Practically speaking, the orbit $(\psi(t),\dot{\psi}(t))$ in the contracting phase, that is with $\tilde{H}<0$, under the transformation (\ref{ttransfrom}), the orbit is the trajectory $(\psi(-t),\dot{\psi}(-t))$ in the expanding phase $\tilde{H}>0$. The energy density $\tilde{\rho}$ in the Einstein frame is given by,
\begin{equation}\label{energydensityeinframe}
\tilde{\rho}=\frac{\dot{\psi}^2}{2\psi^2}+\frac{1}{2k^2\psi^2}\left ( (\delta A_4)^{\frac{1}{1-\delta }}\psi^{n+1}-A_4(\delta A_4)^{\frac{\delta}{1-\delta }}\psi^{n}\right )
\end{equation}
Having in mind that the Hubble parameter $\tilde{H}$ is related to the energy-density $\tilde{\rho}$ by the holonomy corrected FRW equation (\ref{einframeholcorr}), the Hubble parameter vanishes at the point $(\psi,\dot{\psi})=(\delta,0)$ and also at the curve $\tilde{\rho}=\tilde{\rho}_c$. In a future work we shall further study these two different autonomous dynamical systems given by equation (\ref{scalrevol}) implementing the analytic results with a detailed numerical study. In this article we confine ourselves to just qualitatively describe the evolution of the universe in the Einstein frame, and the evolution goes as follows: The universe starts off in the contracting phase $\tilde{H}<0$, oscillates around the unique critical point $(\delta,0)$, then the amplitude of oscillations increases up to the point it reaches the curve $\tilde{\rho}=\tilde{\rho}_c$, at which point the Hubble parameter vanishes. After that, the universe bounces off entering in the expanding phase with $\tilde{H}>0$, expanding in an oscillating way, until it reaches the critical point $(\delta,0)$. We aim to address this LQC corrected $F(R)$ gravity in the Einstein frame, with $F(R)$ being any of the very well known viable gravities, in a future publication. However let us note that the general form of the $\tilde{\rho}=\tilde{\rho}_c$ curve is not as simple as in the $R^2$ gravity case, so a numerical study is compelling.

\section*{Conclusions}

Modified by the attributes of the matter bounce scenario in holonomy corrected LQC, we addressed the question which $F(R)$ gravity can produce the cosmological solutions of matter bounce LQC. Using the reconstruction technique without the need of any auxiliary field, we were able to construct a general differential equation that leads to the matter bounce cosmological solutions. As we demonstrated, this differential equation is a generalized Heun second order differential equation, which in the presence of matter fluids, apart from the pure $F(R)$ geometric fluid, is a non-homogeneous differential equation with polynomial coefficients. We explicitly showed that this equation does not have any polynomial solution and in order to focus on two particular regimes of current experimental interest, late time and early time, we studied in detail the limiting behaviors of the cosmological solutions in these two limits. In the large cosmic time regime, when no matter fluid is present, the $F(R)$ gravity that produces the matter bounce LQC cosmology is an $F(R)$ function with positive rational numbers powers of the Ricci scalar. Interestingly enough, when non-interacting non-relativistic matter is taken into account, the $F(R)$ function that produces the matter bounce cosmology is described by Einstein gravity plus a positive rational power of the Ricci scalar. Note that the large time regime describes eras with high redshift, such as matter domination era or the late time acceleration era, in which case the curvature $R$ takes small values. In the case of small cosmic time, and in the absence of a matter fluid, the $F(R)$ gravity that reproduces the matter bounce solutions is given by a Gauss hypergeometric function, which can be further approximated to positive powers of the Ricci scalar. Notice that the small cosmic time limit corresponds to the early time and also positive powers of the Ricci scalar during this early time regime is, in principle, a favorable feature, since such $F(R)$ functions can consistently describe inflation. In the presence of a relativistic matter fluid, the problem becomes quite complicated, so by using appropriate simplifications by means of asymptotic expansions which can be found in Appendix B, we obtained an $F(R)$ function described by a combination of positive powers of the Ricci scalar. In order to make contact with the experimental data and see to which extend such $F(R)$ gravities can describe consistently inflation, we studied what results are produced by the $F(R)$ gravity we obtained in the small cosmic $t$ limit, without the presence of a matter fluid. In order to analytically obtain the corresponding Einstein frame scalar-tensor theory in terms of the inflaton field, we kept only the most dominating term from the whole $F(R)$ function and this resulted to a spectra index value different from the one obtained by Planck experiment, but of the same order. However, the $r$ index is by far off the predicted value and we believe that in order correct results are obtained, the whole $F(R)$ function should be used. This however would make the analytical solution of the canonical transformation that connects the Jordan and Einstein frame, with respect to $R$, a rather formidable task. In order to implement the $F(R)$ theory predictions on inflation data, we embedded the $F(R)$ theory in a LQC framework, by extending the Hamiltonian constraint in the Einstein frame. The resulting dynamical equations gave similar results to the $R^2$ gravity and the inflation slow roll parameters of the $F(R)$ theory in the Einstein frame can be corrected to some extend in order to have concordance with the experimental data.

It's worth studying the reconstruction of the $F(R)$ theory that produces the matter bounce LQC solutions, using the auxiliary field technique, an issue that is currently under study from us and we hope to report the results soon.

\section*{Acknowledgments}

This work was partially supported by MINECO (Spain) project FIS2010-15640 (S.D.O.)

\section*{APPENDIX A}

In this appendix we shall provide the detailed differential equation that the reconstruction method for the LQC bounce solutions results to, taking also into account the matter fluids contribution. We refer to the differential equation (\ref{bigdiffgeneral1}) appearing in the text. The detailed version of the differential equation is:
\begin{align}\label{expandedfdiff}
& \left( -15552 a^2 A^4 b^2 x^2-3888 a A^3 b^2 x^3+432 A^2 b^2 x^4 \right )\frac{\mathrm{d}^2F(x)}{\mathrm{d}x^2}
\\ \notag & +\Big{(}\frac{9}{2} \sqrt{3} a^2 A^2-\frac{81}{2} \sqrt{3} a^2 A^3+27 \sqrt{3} a^2 A^4+3 \sqrt{3} a A x-27 \sqrt{3} a A^2 x
\\ \notag & +9 \sqrt{3} a A^3 x+\frac{\sqrt{3} x^2}{2}-\frac{9}{2} \sqrt{3} A x^2+5184 a A^3 b^2 x^2-432 A^2 b^2 x^3 \Big{)}\frac{\mathrm{d}F(x)}{\mathrm{d}x}
\\ \notag & -\left ( 9 a^2 A^2+6 a A x+x^2\right )\frac{F(x)}{2}+\sum_iB_{i1}x^{1+w_i}\left ( x+3aA\right )^2=0
\end{align}

\section*{APPENDIX B}

Here we shall present all the approximations we made in order to obtain the $F(R)$ modified that generates the LQC bounce solutions in the presence of a relativistic matter fluid, within the small $t$ (large $R$) approximation. We start off by presenting all the relevant quantities to the calculation of the integral (\ref{finalieseqn}). 

\medskip 

The first expression we shall present is
\begin{align}\label{first}
& \frac{G(z)}{P(z)y_1(z)}\left (y_1(z)^{-2}+(1-z)^{-(A-B)}+z^{-B} \right )=
\\ \notag & \frac{z^{\alpha +\beta } \left(-2 \left(B_1 \left(-\lambda _1+\lambda _2\right)^{4/3}\right) \left(\frac{1}{z}\right)^{2/3}\right)}{z^{\beta } \left(\frac{ \Gamma(-\alpha +\beta ) \Gamma(\gamma )}{\Gamma(\beta ) \Gamma(-\alpha +\gamma )}+\frac{ \alpha  (1+\alpha -\gamma ) \Gamma(-\alpha +\beta ) \Gamma(\gamma )}{(1+\alpha -\beta ) \Gamma(\beta ) \Gamma(-\alpha +\gamma ) z}\right)+z^{\alpha } \left(\frac{ \Gamma(\alpha -\beta ) \Gamma(\gamma )}{\Gamma(\alpha ) \Gamma(-\beta +\gamma )}+\frac{\beta  (1+\beta -\gamma ) \Gamma(\alpha -\beta ) \Gamma(\gamma )}{(1-\alpha +\beta ) \Gamma(\alpha ) \Gamma(-\beta +\gamma ) z}\right)}
\end{align}
Secondly, we made the following approximation, keeping the dominant terms as $z$ tends to infinity,
\begin{align}\label{bigapprox}
&\left (y_1(z)^{2}+(1-z)^{(-A+B)}+z^{B} \right )=\\ \notag &
(1-z)^{-A+B}+z^B+z^{-2 \alpha } \left(\frac{\Gamma(-\alpha +\beta )^2 \Gamma(\gamma )^2}{\Gamma(\beta )^2 \Gamma(-\alpha +\gamma )^2}\right)+z^{-2 \beta } \left(\frac{ \Gamma(\alpha -\beta )^2 \Gamma(\gamma )^2}{\Gamma(\alpha )^2 \Gamma(-\beta +\gamma )^2}\right)\\ \notag & +z^{-\alpha -\beta } \left(\frac{2 \Gamma(\alpha -\beta ) \Gamma(-\alpha +\beta ) \Gamma(\gamma )^2}{\Gamma(\alpha ) \Gamma(\beta ) \Gamma(-\alpha +\gamma ) \Gamma(-\beta +\gamma )}\right)
\end{align}
Finally, the final expression for $u(z)$ is easily obtained by using the above two expressions, namely relations (\ref{first}) and (\ref{bigapprox}), and integrating over $z$
\begin{align}\label{secondhuge}
& u(z)=\frac{12B_1 z^{\frac{1}{3}-B+2 \alpha -3 \beta } \Gamma(\alpha -\beta )^3 \Gamma(\beta ) \Gamma(\gamma )^2 \Gamma(-\alpha +\gamma ) \lambda _1 \left(-\lambda _1+\lambda _2\right)^{1/3}}{(1+3 \alpha -6 \beta ) \Gamma(\alpha )^3 \Gamma(-\alpha +\beta ) \Gamma(-\beta +\gamma )^3} \\ \notag & +\frac{24 B_1 z^{\frac{1}{3}-B+\alpha -2 \beta } \Gamma(\alpha -\beta )^2 \Gamma(\gamma )^2 \lambda _1 \left(-\lambda _1+\lambda _2\right)^{1/3}}{(1-3 \beta ) \Gamma(\alpha )^2 \Gamma(-\beta +\gamma )^2}
\\ \notag & +\frac{12B_1 z^{\frac{1}{3}-B-\beta } \Gamma(\alpha -\beta ) \Gamma(-\alpha +\beta ) \Gamma(\gamma )^2 \lambda _1 \left(-\lambda _1+\lambda _2\right)^{1/3}}{(1-3 \alpha ) \Gamma(\alpha ) \Gamma(\beta ) \Gamma(-\alpha +\gamma ) \Gamma(-\beta +\gamma )}
\\ \notag & +\frac{12B_1 z^{\frac{1}{3}+2 \alpha -\beta } \Gamma(\alpha -\beta ) \Gamma(\beta ) \Gamma(-\alpha +\gamma ) \lambda _1 \left(-\lambda _1+\lambda _2\right)^{1/3}}{(1+3 B+3 \alpha ) \Gamma(\alpha ) \Gamma(-\alpha +\beta ) \Gamma(-\beta +\gamma )}
\\ \notag & +\frac{12B_1 z^{\frac{1}{3}-A+2 \alpha -\beta } \Gamma\left(\frac{4}{3}+\alpha \right) \Gamma(\alpha -\beta ) \Gamma(\beta ) \Gamma(-\alpha +\gamma ) \lambda _1 \left(-\lambda _1+\lambda _2\right)^{1/3}}{\left(\frac{1}{3}-A+B+\alpha \right) (1+3 \alpha ) \Gamma(\alpha ) \Gamma\left(\frac{1}{3}+\alpha \right) \Gamma(-\alpha +\beta ) \Gamma(-\beta +\gamma )}
\\ \notag & -\frac{12B_1 z^{\frac{1}{3}-B+2 \alpha -3 \beta } \Gamma(\alpha -\beta )^3 \Gamma(\beta ) \Gamma(\gamma )^2 \Gamma(-\alpha +\gamma ) \lambda _2 \left(-\lambda _1+\lambda _2\right)^{1/3}}{(1+3 \alpha -6 \beta ) \Gamma(\alpha )^3 \Gamma(-\alpha +\beta ) \Gamma(-\beta +\gamma )^3}
\\ \notag & -\frac{24B_1 z^{\frac{1}{3}-B+\alpha -2 \beta } \Gamma(\alpha -\beta )^2 \Gamma(\gamma )^2 \lambda _2 \left(-\lambda _1+\lambda _2\right)^{1/3}}{(1-3 \beta ) \Gamma(\alpha )^2 \Gamma(-\beta +\gamma )^2}
\\ \notag & -\frac{12B_1 z^{\frac{1}{3}-B-\beta } \Gamma(\alpha -\beta ) \Gamma(-\alpha +\beta ) \Gamma(\gamma )^2 \lambda _2 \left(-\lambda _1+\lambda _2\right)^{1/3}}{(1-3 \alpha ) \Gamma(\alpha ) \Gamma(\beta ) \Gamma(-\alpha +\gamma ) \Gamma(-\beta +\gamma )}
\\ \notag & -\frac{12B_1 z^{\frac{1}{3}+2 \alpha -\beta } \Gamma(\alpha -\beta ) \Gamma(\beta ) \Gamma(-\alpha +\gamma ) \lambda _2 \left(-\lambda _1+\lambda _2\right)^{1/3}}{(1+3 B+3 \alpha ) \Gamma(\alpha ) \Gamma(-\alpha +\beta ) \Gamma(-\beta +\gamma )}
\\ \notag & -\frac{12B_1 z^{\frac{1}{3}-A+2 \alpha -\beta } \Gamma\left(\frac{4}{3}+\alpha \right) \Gamma(\alpha -\beta ) \Gamma(\beta ) \Gamma(-\alpha +\gamma ) \lambda _2 \left(-\lambda _1+\lambda _2\right)^{1/3}}{\left(\frac{1}{3}-A+B+\alpha \right) (1+3 \alpha ) \Gamma(\alpha ) \Gamma\left(\frac{1}{3}+\alpha \right) \Gamma(-\alpha +\beta ) \Gamma(-\beta +\gamma )}
\end{align}
The final expression for $y_2(z)$ is easily obtained by using the approximation (\ref{secondhuge}), keeping the dominant terms as $z$ tends to infinity and integrating over $z$.


\begin{thebibliography}{}




\bibitem{riess} A.G. Riess et al. (High-z Supernova Search Team), Astronom. J. 116, 1009 (1998) [arXiv:astro-ph/9805201]

\bibitem{planck} P.A.R. Ade et al. [arXiv:1302.5082]

\bibitem{bicep} P. A. R. Ade et al. [arXiv:1403.3985], (2014)

\bibitem{reviews1} S. Nojiri, S. D. Odintsov, Int.J.Geom.Meth.Mod.Phys. 11 (2014) 1460006 [arXiv:1306.4426]; Int. J. Geom. Meth. Mod.Phys. 4 (2007) 115 [hep-th/0601213]
 

\bibitem{reviews2} S. Capozziello, V. Faraoni, Beyond Einstein Gravity, Springer, Berlin 2010

\bibitem{reviews3} A. de la Cruz-Dombriz, D. Saez-Gomez, Entropy 14 (2012) 1717 [arXiv:1207.2663]; F. S. N. Lobo, Dark Energy-Current Advances and Ideas, 173-204 (2009) [arXiv:0807.1640]

\bibitem{reviews4} S. Nojiri, S. D. Odintsov,  Phys.Rept. 505 (2011) 59 [arXiv:1011.0544]

\bibitem{reviews5} S. Capozziello, M. De Laurentis, Phys.Rept. 509 (2011) 167 [arXiv:1108.6266]


\bibitem{reviews8} K. Bamba, S. Nojiri, S. D. Odintsov, JCAP 0810 (2008) 045 [arXiv:0807.2575]

\bibitem{reviews9} S. Nojiri, S. D. Odintsov, Phys.Lett. B657 (2007) 238 [arXiv:0707.1941];  Gen.Rel.Grav. 36 (2004) 1765 [hep-th/0308176] 



\bibitem{importantpapers1} S. Capozziello, S. Nojiri, S.D. Odintsov, A. Troisi, Phys.Lett. B639 (2006) 135 [astro-ph/0604431]; S. Nojiri, S. D. Odintsov, Phys.Rev. D77 (2008) 026007 [arXiv:0710.1738]

\bibitem{importantpapers2} S. Capozziello, V.F. Cardone, S. Carloni, A. Troisi, Int.J.Mod.Phys. D12 (2003) 1969 [astro-ph/0307018]

\bibitem{importantpapers3} S. Nojiri, S. D. Odintsov, Phys.Rev. D74 (2006) 086005 [hep-th/0608008]; S. Nojiri, S.D. Odintsov, D. Saez-Gomez, AIP Conf.Proc. 1458 (2011) 207 [arXiv:1108.0767]; S. Nojiri, S. D. Odintsov, J.Phys.Conf.Ser. 66 (2007) 012005 [hep-th/0611071]; A. de la Cruz-Dombriz, A. Dobado, Phys.Rev. D74 (2006) 087501 [gr-qc/0607118] 

\bibitem{importantpapers4} W. Hu, I. Sawicki, Phys.Rev.D76 (2007) 064004 [arXiv:0705.1158] 

\bibitem{importantpapers5} S. M. Carroll, V. Duvvuri, M. Trodden, M. S. Turner, Phys.Rev. D70 (2004) 043528 [astro-ph/0306438]; S. Capozziello, Int.J.Mod.Phys.D11, 483 (2002) [gr-qc/0201033]

\bibitem{importantpapers6} R. Myrzakulov, L. Sebastiani, S. Zerbini, Int.J.Mod.Phys. D22 (2013) 1330017 [arXiv:1302.4646]


\bibitem{importantpapers8} O. Bertolami, R. Rosenfeld, Int.J.Mod.Phys. A23 (2008) 4817 [arXiv:0708.1784]

\bibitem{importantpapers9} A. Capolupo, S. Capozziello, G. Vitiello, Int.J.Mod.Phys. A23 (2008) 4979 [arXiv:0705.0319]

\bibitem{importantpapers10}   P. K.S. Dunsby, E. Elizalde, R. Goswami, S. Odintsov, D. S. Gomez, Phys.Rev. D82 (2010) 023519 [arXiv:1005.2205]

\bibitem{importantpapers11} G. Cognola, E. Elizalde, S. Nojiri, S.D. Odintsov, L. Sebastiani, S. Zerbini, Phys.Rev. D77 (2008) 046009 [arXiv:0712.4017 ]; K. Bamba, Chao-Qiang Geng, Chung-Chi Lee, JCAP 1008 (2010) 021 [arXiv:1005.4574]


\bibitem{importantpapers12}   S. Nojiri, S. D. Odintsov, D. Saez-Gomez, Phys.Lett. B681 (2009) 74 [arXiv:0908.1269]  

\bibitem{importantpapers13} S. Capozziello, V. F. Cardone, A. Troisi, Phys.Rev. D71 (2005) 043503 [astro-ph/0501426]

\bibitem{importantpapers14} J. C.C. de Souza, Valerio Faraoni, Class.Quant.Grav. 24 (2007) 3637 [arXiv:0706.1223]; V. Faraoni, Phys.Rev. D74 (2006) 104017 [astro-ph/0610734];G. J. Olmo, Phys.Rev.Lett. 95 (2005) 261102 [gr-qc/0505101]; G. J. Olmo, Phys.Rev. D75 (2007) 023511 [gr-qc/0612047] 

\bibitem{importantpapers15} S. A. Appleby, R. A. Battye, A. A. Starobinsky, JCAP 1006 (2010) 005 [arXiv:0909.1737]
 

\bibitem{importantpapers17} S. A. Appleby, R. A. Battye, Phys.Lett.B654 (2007) 7 [arXiv:0705.3199]; S. A. Appleby, R. A. Battye, JCAP 0805 (2008) 019 [arXiv:0803.1081]

\bibitem{importantpapers18} A. Silvestri, M. Trodden, Rept. Prog. Phys. 72 (2009) 096901 [arXiv:0904.0024]

\bibitem{importantpapers19} E. Elizalde, E.O. Pozdeeva, S.Yu. Vernov, Phys.Rev. D85 (2012) 044002 [arXiv:1110.5806]

\bibitem{importantpapers20} V. Faraoni,  Phys.Rev. D75 (2007) 067302 [gr-qc/0703044]


\bibitem{sergeinojirimodel} S. Nojiri, S. D. Odintsov, Phys.Rev. D68 (2003) 123512 [hep-th/0307288]

\bibitem{capo} M. Sami, Curr. Sci. 97,887(2009) [arXiv:0904.3445]; Yi-Fu Cai, E. N. Saridakis, M. R. Setare, Jun-Qing Xia, Phys.Rept. 493 (2010) 1 [ arXiv:0909.2776]

 \bibitem{capo1} T. Padmanabhan, Phys.Rept. 380 (2003) 235 [hep-th/0212290]; K. Bamba, S. Capozziello, S. Nojiri, S. D. Odintsov, Astrophys. Space Sci. 342, 155 (2012) [arXiv:1205.3421]

\bibitem{peebles} P.J.E. Peebles, Bharat Ratra, Rev.Mod.Phys. 75 (2003) 559 [astro-ph/0207347]; V. Sahni, AIP Conf.Proc. 782 (2005) 166, J.Phys.Conf.Ser. 31 (2006) 115; M. Li, Xiao-Dong Li, S. Wang, Yi Wang, Commun.Theor.Phys. 56 (2011) 525 [arXiv:1103.5870]; A. Joyce, B. Jain, J. Khoury, M. Trodden [arXiv:1407.0059]

\bibitem{faraonquin} V. Faraoni, Int.J.Mod.Phys. D11 (2002) 471 [astro-ph/0110067]; V.K. Onemli, R.P. Woodard, Class.Quant.Grav. 19 (2002) 4607 [gr-qc/0204065] 

\bibitem{tsujiintjd} A. Gomez-Valent, J. Sola, S. Basilakos, arXiv:1409.7048 
 

\bibitem{LQC1}A. Ashtekar, P. Singh, Class. Quant. Grav. 28, 213001 (2011) [arXiv:1108.0893 ]

\bibitem{LQC2} A. Ashtekar, Nuovo Cim. B122 (2007) 135 [gr-qc/0702030]


\bibitem{LQC3} A. Corichi, P. Singh, Phys.Rev. D80 (2009) 044024 [arXiv:0905.4949]

\bibitem{LQC4} P. Singh, K. Vandersloot, G.V. Vereshchagin, Phys.Rev. D74 (2006) 043510 [gr-qc/0606032]

\bibitem{LQC5sing} P. Singh, Class.Quant.Grav. 26 (2009) 125005 [arXiv:0901.2750]

\bibitem{LQC6sing} A. Ashtekar, T. Pawlowski, P. Singh, Phys.Rev. D74 (2006) 084003 [gr-qc/0607039]

\bibitem{LQC7sing} M. Bojowald, Class.Quant.Grav. 26 (2009) 075020 [arXiv:0811.4129]


\bibitem{LQC7sing1} M. Sami, P. Singh, Shinji Tsujikawa, Phys.Rev. D74 (2006) 043514


\bibitem{LQC8} E.J. Copeland, D.J. Mulryne, N.J. Nunes, M. Shaeri, Phys.Rev. D77 (2008) 023510 [arXiv:0708.1261]

\bibitem{LQC9} D. Samart, B. Gumjudpai, Phys.Rev. D76 (2007) 043514 [gr-qc/0605113]

\bibitem{LQC10} T. Naskar, J. Ward, Phys.Rev. D76 (2007) 063514 [arXiv:0704.3606 ]



\bibitem{LQC11} T. Cailleteau, J. Mielczarek, A. Barrau, J. Grain, Class.Quant.Grav. 29 (2012) 095010 [arXiv:1111.3535 ]

\bibitem{LQC12} T. Cailleteau, A. Barrau, J. Grain, F. Vidotto, Phys.Rev. D86 (2012) 087301 [arXiv:1206.6736]

\bibitem{LQC13} T. Cailleteau, A. Barrau, J. Grain, F. Vidotto, Phys.Rev. D86 (2012) 087301 [arXiv:1206.6736]



\bibitem{mbounce1} R. H. Brandenberger [arXiv:1206.4196; J. Quintin, Yi-Fu Cai, R. H. Brandenberger, Phys. Rev. D90 (2014) 063507 [arXiv:1406.6049 ]; Yi-Fu Cai, D. A. Easson, R. Brandenberger, JCAP 1208 (2012) 020 [arXiv:1206.2382] ; Yi-Fu Cai, R. Brandenberger, X. Zhang, Phys.Lett. B703 (2011) 25 [arXiv:1105.4286] 

\bibitem{mbounce2} Yi-Fu Cai, R. Brandenberger, X. Zhang, JCAP 1103 (2011) 003 [arXiv:1101.0822]; C. Li, R. H. Brandenberger, Yeuk-Kwan E. Cheung [arXiv:1403.5625; Yi-Fu Cai, E. McDonough, F. Duplessis, R. H. Brandenberger, JCAP 1310 (2013) 024 [arXiv:1305.5259]; R. H. Brandenberger [arXiv:1206.4196  


\bibitem{mbounce3} P. Singh, Class.Quant.Grav. 26 (2009) 125005 [arXiv:0901.2750]


\bibitem{mbounce4} J. Amoros, J. Haro, S. D. Odintsov, Phys.Rev. D87 (2013) 104037 [arXiv:1305.2344]; T. Qiu, X. Gao, E. N. Saridakis, Phys.Rev. D88 (2013) 4, 043525 [arXiv:1303.2372]


\bibitem{mbounce5} J. Haro, Europhys. Lett. 107 (2014) 29001 [arXiv:1403.4529]


\bibitem{mbounce6} Yi-Fu Cai, Shih-Hung Chen, J. B. Dent, S. Dutta, E. N. Saridakis, Class.Quant.Grav. 28 (2011) 215011 [arXiv:1104.4349]


\bibitem{mbouncersquarefr} J. Amoros, J. de Haro, S.D. Odintsov, Phys.Rev. D89 (2014) 104010 [arXiv:1402.3071]


\bibitem{mbounce8} E. Wilson-Ewing, JCAP 1303 (2013) 026 [arXiv:1211.6269]


\bibitem{mbounce9} Yi-Fu Cai, E. Wilson-Ewing, JCAP 1403 (2014) 026 [arXiv:1402.3009 ]

\bibitem{mbounce10} J. Haro, J. Amoros [arXiv:1406.0369

\bibitem{mbounce11} J. Haro, J. Amoros, JCAP 08(2014)025 [arXiv:1403.6396 ]

\bibitem{kaiser} D. I. Kaiser, Phys.Lett. B340 (1994) 23 [astro-ph/9405029]; D. I. Kaiser, Phys.Rev. D52 (1995) 4295 [astro-ph/9408044] 

\bibitem{cairev3} Yi-Fu Cai, Shih-Hung Chen, J. B. Dent, S. Dutta, E. N. Saridakis, Class. Quantum Grav. 28 (2011) 215011 [arXiv:1104.4349]

\bibitem{cairev1} Yi-Fu Cai, Sci.China Phys.Mech.Astron. 57 (2014) 1414 [arXiv:1405.1369]

\bibitem{sergeibounce} K. Bamba, A. N. Makarenko, A. N. Myagky, S. Nojiri, S. D. Odintsov, JCAP01(2014)008  [arXiv:1309.3748]

\bibitem{sergeiinflnew} K. Bamba, S. Nojiri, S. D. Odintsov, D. Saez-Gómez, arXiv:1410.3993 

\bibitem{mathpaper} Yao-Zhong Zhang, J. Phys. A: Math. Theor. 45 (2012) 065206 [arXiv:1107.5090]

\bibitem{khoury} J. Khoury, A. Weltman, Phys. Rev. D69, 044026 (2004) [astro-ph/0309411 ]

\bibitem{sergeistarobinsky} L. Sebastiani, G. Cognola, R. Myrzakulov, S.D. Odintsov, S. Zerbini, Phys. Rev. D 89, 023518 (2014) [arXiv:1311.0744]

\bibitem{holonomyfr} X. Zhang, Y. Ma, Phys.Rev.Lett. 106, 171301 (2011) [arXiv:1101.1752]

\bibitem{mukhanov} V. Mukhanov, Physical foundations of cosmology, Cambridge, UK: Univ. Pr. (2005) 421 p; D. S. Gorbunov, V. A. Rubakov, Introduction to the theory of the early universe: Cosmological perturbations and inflationary theory, Hackensack, USA, World Scientific (2011) 489 p

\bibitem{guth} A. H. Guth, Phys.Rept. 333 (2000) 555 [astro-ph/0002156]; A. H. Guth,  J.Phys. A40 (2007) 6811 [hep-th/0702178]

\bibitem{starobinsky} A. A. Starobinsky, Phys.Lett. B91 (1980) 99

\bibitem{sergeitrace} K. Bamba, R. Myrzakulov, S. D. Odintsov, L. Sebastiani, Phys. Rev. D 90, 043505 (2014) [arXiv:1403.6649]






\end{thebibliography}
\end{document}